# Small Angle Scattering and Zeta Potential of Liposomes Loaded with Octa(carboranyl)porphyrazine

*Anna Salvati,[a] Sandra Ristori,[a] Julian Oberdisse,[b] Olivier Spalla,[c] Giampaolo Ricciardi,[d] Daniela Pietrangeli,[d] Mauro Giustini,[e] Giacomo Martini [a*].*

a) University of Firenze, Department of Chemistry, Via della Lastruccia 3, 50019 Sesto Fiorentino, Italy.

martini@csgi.unifi.it

* Corresponding author: Giacomo Martini

E-mail: martini@csgi.unifi.it

Telephone number: 00390554573042

Fax number: 00390554573036

b) Laboratoire des Colloïdes, Verres et Nanomatériaux (LCVN), Université Montpellier II, 34095 Montpellier Cedex 05 France and Laboratoire Léon Brillouin, CEA Saclay, 91191 Gif sur Yvette, France

c) CEA Saclay, DRECAM-SCM-LIONS, 91191 Gif Sur Yvette, France

d) University of Basilicata, Department of Chemistry, Via N. Sauro 85, 85100 Potenza, Italy

e) University "La Sapienza", Department of Chemistry, P.le Aldo Moro 5, 00185 Roma, Italy

ABSTRACT In this work the physicochemical characterization of liposomes loaded with a newly synthesised carboranyl porphyrazine ($H_2HECASPz$) is described. This molecule represents a potential drug for different anticancer therapies, such as Boron Neutron Capture Therapy, Photodynamic Therapy and Photothermal Therapy. Different loading methods and different lipid mixtures were tested. The corresponding loaded vectors were studied by Small Angle Scattering (SANS and SAXS), light scattering and zeta potential. The combined analysis of structural data at various length scales and the measurement of the surface charge allowed to obtain a detailed characterization of the investigated systems. The mechanisms underlying the onset of differences in relevant physicochemical parameters (size, polydispersity and charge) were also critically discussed.

INTRODUCTION

Drug carriers are usually required in pharmaceutical formulations for several reasons: first, most of therapeutic agents are scarcely soluble in aqueous environments, such as the body fluids or the cell interior; secondly, many drugs, as well as proteins or DNA, need to be protected from undesired interactions which could yield hydrolysis, enzymatic degradation or loss of the native structure and, consequently, of their activity. Finally, drugs may be toxic and have to be compartmentalized until they reach the target tissue. At the same time, carriers can be engineered to include surface functionalities, such as targeting ligands, to improve the selectivity of delivery.

Moreover, both vectors and their load may undergo chemical and physical modifications, that affect the structure of the system, its loading and release capacity, interfacial properties, and, eventually, biodistribution [1, 2]. In some cases, the drug itself triggers these modifications, and may also destabilize the starting system. Therefore a physical and chemical characterization of loaded vector is necessary, to fully understand the biological efficiency.

Liposomes are versatile carriers formed by molecules that self-assemble and were first proposed as drug delivery vehicles by Gregoriadis more than 30 years ago [3]. Since then, liposomes have been extensively used for various delivery and imaging applications, including commercially available products [4-8].

In this work, different kinds of liposomes have been chosen to solubilise a highly hydrophobic porphyrazine derivative (2,3,7,8,12,13,17,18-octakis-(1-allyl-1,2-dicarba-closododecaboran-2-yl)-hexylthio-5,10,15,20-(21H,23H)Porphyrazine, henceforth called H$_2$HECASPz) which contains eight carborane icosahedral cages. The synthesis of this molecule and its properties have been previously described [9]. This molecule represents a potential drug for Boron Neutron Capture Therapy, BNCT, and as photosensitizers for Photodynamic Therapy, PDT, or Photothermal Therapy, PTT. In this work, two-component mixed liposomes were built up with 1,2-dioleoyl-*sn*-glycero-phosphocholine (DOPC), 1,2-dioleoyl-*sn*-glycerophosphoethanolamine (DOPE), 1,2-dioleoyl-3-trimethylammoniuum-propane chloride (DOTAP), and 1,2-dioleoyl-*sn*-glycero-3-phosphate monosodium salt (DOPA) (Scheme 1). In all cases, long and flexible oleoyl chains were chosen to accommodate the bulky porphyrazine. Indeed, the distance between two sulphur atoms across the porphyrazine core is 2 nm and the size of the whole molecule in its extended form is 4 nm, as determined by DFT calculations. This latter value is close to the bilayer thickness of dioleoyl phospholipids (4-5 nm, depending on the polar head type) which, joint to the well known rigidity of porphyrin-like macrocycles, might hamper insertion. However, for the H$_2$HECASPz used in the present work a specific interaction between the phospholipid heads and the heterocyclic ring has been postulated on the basis of spectrophotometric data and theoretical modelling[9]. Such interaction may compensate for predictable unfavourable factors. The model proposed to describe the loading modality of H$_2$HECASPz into liposomes features the porphyrazine ring right above the polar head of lipids and the hydrocarbon plus boron cage substituents interwoven among the oleoylic chains.

DOPE was added as helper lipid in all formulations because of its well documented ability to form columnar inverted hexagonal liquid-crystalline structures, which improve the fusogenic properties of lipidic vectors [10-12]. Positive, negative, and zwitterionic phospholipids were therefore used with the aim of testing the effect of different polar heads and surface charge in drug loading [13]. They were characterized by means of Small Angle X-ray Scattering, Small Angle Neutron Scattering (SANS), Light Scattering (LS), and Zeta Potential (ZP) to obtain a detailed picture of the aggregates prior to their use in biological tests.

Porphyrazine-loaded liposomes were prepared according to the detergent depletion method [14] (henceforth called "detergent depletion liposomes", DDL) with sodium cholate or tetradecyltrimethylammonium bromide ($C_{14}TAB$) as helping surfactant, in order to improve the incorporation of the hydrophobic carboranylporphyrazine within lipid bilayers. The properties of the liposomes obtained by this method were compared with those of liposomes prepared by the more commonly used extrusion procedure (henceforth called "extruded liposomes", EL). This comparison revealed that marked differences can be obtained in the structural features of the same liposome formulation prepared according to different methods and that porphyrazine content was invariably 3-5 times higher in liposomes obtained by detergent depletion.

MATERIAL AND METHODS

DOTAP (purity >99%, *1*), DOPA (purity >99%, *2*), and DOPE (purity >99%, *3*) were purchased from Avanti Polar Lipids, Inc., Alabaster, AL. DOPC (purity >99%, *4*) was purchased from Northern Lipids, Inc., Vancouver, Canada. All lipids were used as received. The 2,3,7,8,12,13,17,18-octakis-(1,2-dicarba-*closo*-dodecaboranyl)-hexylthio-5,10,15,20-porphyrazine $H_2HECASPz$, *5*, was synthesized as described in ref.[9] Sodium cholate, *6*, and Sephadex G-50 superfine were purchased from Sigma Aldrich, and tetradecyl trimethylammonium bromide ($C_{14}TAB$, *7*) was from Fluka.

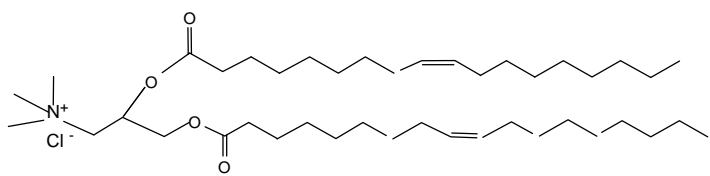

*1*

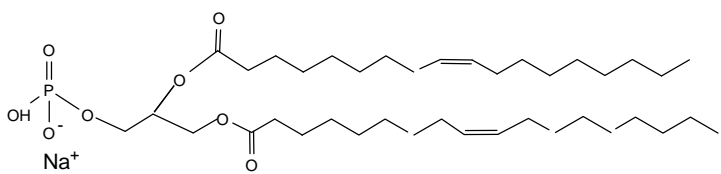

*2*

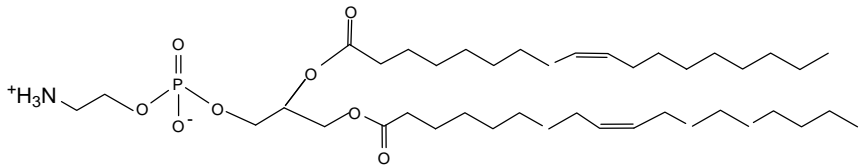

*3*

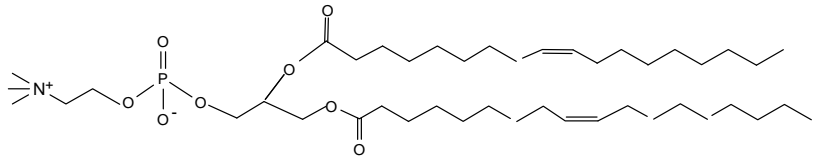

*4*

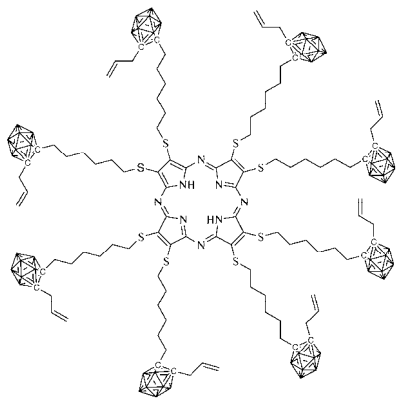

*5*

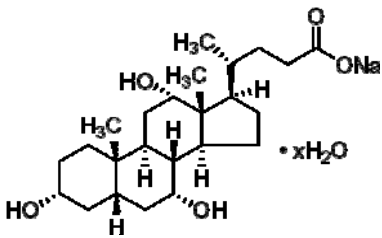

*6*

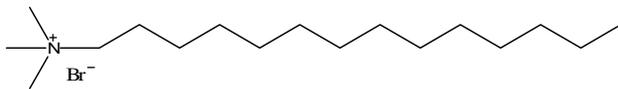

*7*

*Liposome preparation:* DOTAP/DOPE, DOPC/DOPE, DOPA/DOPE liposomes at 1:1 lipid mole ratio and pure DOPC, with a $1 \cdot 10^{-2}$ mol/L total lipid concentration were prepared according two different methods, *viz* detergent depletion (DDL) and extrusion (EL). For each formulation, three different samples were prepared: 1) pure liposome; 2) liposome containing $0.5 \cdot 10^{-4}$ mol/L H$_2$HECASPz (porphyrazine/lipid ratio = 1/200); 3) liposome containing $1 \cdot 10^{-4}$ mol/L H$_2$HECASPz (porphyrazine/lipid ratio = 1/100). Samples for SANS experiments were prepared in D$_2$O in order to enhance contrast between the solvent and the aggregates.

For DDL, sodium cholate was chosen in the case of DOPC/DOPE, DOPC and DOPA/DOPE liposomes, whereas C$_{14}$TAB was used for DOTAP/DOPE liposomes, since in this case the negatively charged helper surfactant strongly interacted with cationic lipids and resulted in phase separation. In order to compare the effect of different detergents on the same formulation, DOPC/DOPE liposomes were prepared by using both sodium cholate and C$_{14}$TAB. The mixed films were hydrated with a highly concentrated solution of the chosen detergent (3:1 detergent:lipid molar ratio, i.e. 3x10$^{-2}$ M starting detergent concentration). Samples were then sonicated till clarity, to form mixed micelles and the detergent was removed by eluting the suspension through a Sephadex G-50 column with the same solvent [14, 15].

ELs were prepared by hydrating the lipid-carboporphyrazine mixed film with pure water, followed by 8 freeze-thaw cycles and 27 passages through 200 nm pore size membranes, as described in ref. [16]. The more common 100 nm membranes were not used because of the rigidity imparted to the lipid film by the included porphyrazine molecules.

H$_2$HECASPz incorporation rate was measured spectrophotometrically after liposome disruption. In a typical experiment, the liposome solution (500 μL) was freezed overnight at 253 K, warmed at 353 K in a water bath for 2h, then poured on anhydrous Na$_2$SO$_4$ (2.0 g). The deliquescent solid mixture was then extracted with chloroform (3 times with 15 mL each). The organic extracts were collected and the volume reduced to 2.0 mL. The porphyrazine concentration was finally monitored by measuring the absorbance at 712 nm (Q band, molar absorption coefficient $\varepsilon = 2.8 \times 10^4$ M$^{-1}$ cm$^{-1}$), after subtraction of the blank, i.e. the spectrum of plain liposomes treated in the same way as porphyrazine loaded ones.

*Zeta potential* (ζ). The measurements were performed with a Coulter DELSA 440 SX (Coulter Corporation, Miami, FL, USA). ζ was calculated from the electrophoretic mobility by means of the Helmholtz-Smoluchowski relationship. More details are reported in ref. [16]. All samples were diluted 15 times to meet the instrumental sensitivity requirements.

*Size measurements.* The liposome size was measured by dynamic light scattering (DLS) with a Coulter Sub-Micron Particle Analyzer, Model N4SD, equipped with a 4 mW helium-neon laser (632.8 nm) and 90° detector. The autocorrelation function of the scattered light was analyzed by unimodal analysis, that assumes a log Gaussian distribution of the vesicle size [17] and by SDP (Size Distribution Processor), based on the algorithm CONTIN [18]. This allowed to obtain mean diameter and polydispersity index, $(\mu/\gamma^2)^2$, of plain- and drug-loaded liposomes.

*SANS measurements.* SANS experiments were performed at the Laboratoire Léon Brillouin (Saclay, France) by collecting data in three different configurations ($\lambda = 12$ Å and D = 5 m; $\lambda = 4$ Å and D = 5 m; $\lambda = 4$ Å and D = 1 m, where $\lambda$ is the neutron wavelength and D the sample-detector distance). The total $q$-range covered was 0.0047-0.54 Å$^{-1}$. SANS profiles were reduced to the absolute scale and the intensity at large angles (incoherent background) was subtracted before fitting the data. All the fits were carried out using the form factor of polydisperse vesicles and including the instrumental resolution function.

Contrarily to the SAXS analysis, which concentrated on the bilayer internal structure, the lower-$q$ data of the SANS investigation allowed to capture the $q^{-2}$ power law, characteristic of bilayer scattering. The complete form factor of monolamellar vesicles was used in the modeling:

$$P_{ves} = F_{shell}^2(q, R_n) = \left\{ \frac{4\pi}{3}(R_n + t_n)^3 f\left[q(R_n + t_n)\right] - \frac{4\pi}{3} R_n^3 f(qR_n) \right\}^2 \qquad [1a]$$

$$f(x) = \frac{\sin x - x\cos x}{x^3} \qquad [1b]$$

where $R_n$ is the inner radius in the vesicle and $t_n$ is total bilayer thickness measured by SANS. In the absence of any intervesicle interaction, the scattered intensity I(q) then reads:

$$I(q) = \frac{n}{V} \Delta\rho^2 P_{ves} \qquad [2]$$

where Δρ is the neutron scattering length density difference between the vesicle and the solvent (SANS contrast), $n/V$ is the number density of scattering objects, i.e. the number of vesicles per unit volume. All fits were carried out by using the form factor in eq. [1], integrated numerically over a Gaussian distribution function in radius (with standard deviation $\sigma_R$), in thickness (with standard deviation $\sigma_{tn}$), and convoluted with the resolution function of the instrument [19, 20]. Note that X-ray and neutron scattering probe different parts of the lipid bilayer. Indeed, X-ray scattering is more sensitive to the polar head region where the electronic density is higher than in the bilayer core. On the contrary, neutrons "see" the distribution of protons in the $D_2O$, with some contribution from oxygen and carbon atoms [21]. It has been calculated that in the case of hydrogenated glycerophospholipid vesicles in $D_2O$, the maximum of the scattering length density for neutrons is located at the level of the carbonyl groups [22, 23], whereas for X-ray the major contribution comes from the phosphate units. This should account for a few Ångstrom difference in the bilayer length obtained by the two techniques.

*SAXS measurements.* SAXS experiments were performed at the ID2 beamline of the ESRF (European Synchrotron Radiation Facility, Grenoble, France) [24]. The energy of the incoming beam was 12 keV and the wavelength was 1.0 Å. Samples were placed in 1.5 mm diameter capillaries. Two sample-detector distances were used (1 m and 3 m), leading to a total *q*-range of 0.01-0.534 Å$^{-1}$. The profiles of the DOTAP/DOPE detergent depletion series were recorded with a different detector, allowing to cover a comparable *q* range (0.0142-0.420 Å$^{-1}$) with only one sample-detector distance (1 m). Calibration to the absolute scale of the experimental curves was performed according to a well established procedure [25]. Briefly, the raw data were first radially averaged, divided by the acquisition time and by transmission; then background correction was performed by subtracting the empty holder SAXS curve, prior to thickness normalisation. Finally, the experimental curve of a reference water-filled capillary subjected to the same mathematical treatment was subtracted. The equation of a bilayer form factor was used to fit the scattered intensity:

$$I(q) = \frac{4\pi}{q^4} \Sigma \left[ (\rho_{c,e} - \rho_{h,e}) \sin(q\frac{t_c}{2}) + (\rho_{h,e} - \rho_{s,e}) \sin(q\frac{2t_h + t_c}{2}) \right]^2 \qquad [3]$$

where $\rho_c$, $\rho_h$, $t_c$ and $t_h$ are the electronic density and the thickness of the hydrophobic and hydrophilic layer, respectively; $\rho_{s,e}$ is the electronic density of the solvent and $\Sigma$ the interface extension per unit volume [26, 27]. A Gaussian distribution of the bilayer hydrophobic thickness and, when necessary, of the polar head thickness was used to take polydispersity into account (half-width at half maximum $\sigma_{t_c}$ and $\sigma_{t_h}$, respectively).

The volume fraction of liposomes was obtained from the concentration and average density of the lipid molecules. Then, assuming a thickness of the hydrophilic and hydrophobic part, without taking the host porphyrazines into account (since in absolute terms they amount to less than 1% of all molecules), we calculated a total specific surface $\Sigma$. This value was inserted in the prefactor of the SAXS fitting expression. Iterative refinement was done on the basis of thickness values extracted from SAXS itself, until a satisfactory agreement was reached.

RESULTS AND DISCUSSION

1) *DOTAP/DOPE cationic liposomes.*

Figure 1 shows the experimental SANS curves of unloaded and porphyrazine-loaded DOTAP/DOPE liposomes prepared by detergent depletion (on the left) and by extrusion (on the right) and the best fit curves for the corresponding pure liposomes.

The incoherent background (i.e. the hydrogen content) of porphyrazine loaded liposomes, before the subtraction performed as described in the Materials and Methods section, increased, as expected, with increasing H$_2$HECASPz concentration. However, the low signal/noise ratio in the high $q$ SANS region did not allow to quantify the amount of inserted guest molecules.

The minimum at $q$ = 0.16-0.18 Å$^{-1}$ corresponded to the onset of the form factor decay, due to the scattering length density distribution across the bilayer, i.e. to the bilayer form factor, which is proportional to $[1 - \cos(qt)]$ [25] in homogeneous planar bilayers and has a minimum at $2\pi = qt_n$. This minimum, when pronounced, is indicative of monodisperse bilayers. In our case, it turned to an inflection region in the DDL curves and remained rather shallow in the EL curves. This suggested the presence of polydisperse bilayers. For both DDLs and ELs studied in this work no Guinier plateau was found at low $q$ values in the SANS diagrams, indicating that the vesicles were too large for the $q$ range investigated. Thus, the radius measured by DLS was taken as an input parameter in the fitting.

More detailed information about the liposome bilayers were obtained by SAXS. Figure 2 compares experimental and fitted SAXS profiles of plain cationic DDLs and ELs. Table 1 reports the best fit parameters of the SAXS and SANS curves shown in Figures 1 and 2, together with the light scattering and zeta potential results for the cationic DDLs and ELs.

DDLs with $C_{14}$TAB used as detergent had a high polydispersity index, as compared to the same ELs samples. The reason for using a cationic surfactant in these DDLs arose from the strong interaction which occurred between lipids and oppositely charged detergent molecules.

The length of the alkyl chain of TAB surfactant was chosen to reach a compromise among good solubilization properties, critical micellar concentration (cmc), and low Krafft temperature. This compromise was met with $C_{14}$TAB. Shorter chain alkylammonium salts, though characterized by a higher cmc, showed scarce solubilization of the porphyrazine, even after prolonged sonication. This resulted in vesicles suspensions with a low amount of $H_2$HECASPz solubilised in the dispersed lipid/surfactant phase (data not shown). In all cases, no micelles were detected by dynamic light scattering in the final DOTAP/DOPE preparation. This ensured that the samples contained only liposomes. At the same time, the similarity in the zeta potential values of pure liposomes obtained by the two methods of preparation showed that removal of the helper surfactant was achieved. Indeed, the residual content of helper surfactant in DDLs is known to be of the order of a few percent [14, 15, 28].

$H_2$HECASPz content in loaded liposomes was determined spectrophotometrically as described in the Materials and Methods section. The loading rate was 30-40% for DDLs, appreciably higher than 7-12% found for liposomes extruded through 200 nm membranes.

At intermediate $q$, the intensity was proportional to $q^{-2}$, which is the characteristic power law for bilayer scattering. This confirmed that the liposome structure was not disrupted upon $H_2$HECASPz insertion.

The experimental SAXS profiles of porphyrazine-loaded ELs were very close to one another and to that of plain ELs, analogously to what observed for the SANS curves. Therefore, only one set of SAXS best fit parameters is given in Table 1. More marked differences were observed in the structure of DDLs, as given by SAXS profiles, before and after $H_2HECASPz$ insertion (Figure 3).

The fitting of DDL SAXS curves allowed to establish that the detergent depletion procedure gave liposomes with higher polydispersity (in line with the other measurements reported in this work) and with thicker bilayers.

*2) DOPA/DOPE anionic liposomes*

Negatively charged DOPA/DOPE liposomes were prepared by extrusion and by detergent depletion with sodium cholate as helper surfactant. Figure 4 shows the experimental SANS curves of DOPA/DOPE DDLs and the best fit curve for the pure liposomes. The three patterns were very similar to one another. No large differences were either detected by neutron scattering for the three corresponding EL samples. For this reason, the SANS curves are not shown. Table 2 shows DLS data, ζ and best fit parameters for SAXS and SANS profiles of the two series.

Sodium cholate as helper detergent in DDLs gave rise to monodisperse aggregates. The average size was larger than for DOTAP/DOPE DDLs prepared with $C_{14}TAB$. This suggested that the use of a bile salt derivative, such as cholate, was preferable to obtain more homogeneous liposomes, and resulted in a less curved bilayer, i.e. in a larger size of the final objects. ELs had a mean diameter which agreed with the membrane pores and were monodisperse, as expected from this method of preparation. As described above for the cationic liposomes, the zeta potential of pure liposomes prepared by the two procedures gave similar results. This ensured that detergent removal was achieved with the polydextrane column.

$H_2HECASPz$-loading was 25-40% in DDLs and 10-20% in ELs. In agreement with these analytical data, the differences in size and zeta potential at increasing $H_2HECASPz$ content were small in the case of ELs and were more pronounced in DDLs. Zeta value increased with increasing porphyrazine concentration, thus suggesting that $H_2HECASPz$ diluted with its insertion the negative charge of DOPA/DOPE liposomes.

As already found for DLS results and ζ values, DDLs showed large differences upon H$_2$HECASPz insertion. The best fit values listed in Table 2 indicated that progressive insertion of the porphyrazine resulted not only in the increase of aggregate size (in agreement with DLS results) but also of the bilayer thickness and polydispersity. However, the overall effects of the insertion did not affect substantially the structure of liposomes, which remained intact after drug loading.

The SAXS analysis confirmed these observations. Figure 5 shows the SAXS experimental curves of plain and porphyrazine loaded liposomes (lipid/ H$_2$HECASPz = 100/1 starting ratio) prepared by the two methods. Accordingly, the three curves of ELs were fitted with a unique set of parameters, whereas in the case of DDLs it was possible to distinguish among systems obtained from different starting porphyrazine concentration.

*3) DOPC/DOPE zwitterionic liposomes*

Figure 6 shows the experimental SANS profiles of plain and loaded DOPC/DOPE samples prepared by detergent depletion with sodium cholate and C$_{14}$TAB as helper surfactants (left and right, respectively). The fit of the pure liposome scattering curves is also shown in the same figure.

Table 3 reports the DLS data and the best fit parameters of the SANS diagrams for these DDLs.

As outlined in the experimental section, we chose zwitterionic liposomes to investigate possible differences induced by the two helper detergents used in the preparation of liposomes. Interestingly, for DDLs with C$_{14}$TAB the obtained H$_2$HECASPz loading ratio was higher than for all other preparations reported in this work and reached up to 68% of the initial porphyrazine content, while in the case of sodium cholate the loading was 38-55% [9].

The size of DOPC/DOPE liposomes taken from DLS results gave satisfactory fittings of the experimental diagrams for DDLs.

In the case of Els the experimental SANS curves of plain- and porphyrazine-loaded were very similar. The mean diameter obtained by DLS was 1500-1600 Å, and the polydispersity index was 0.05. For plain Els, fitting was done with monolamellar vesicles with $R_n$ = 800 Å, $\sigma_R$ = 20 Å, $t_n$ = 36 Å, and $\sigma_{tn}$ = 2 Å). This indicated substantial invariance after H$_2$HECASPz uptake by co-extrusion, as could be expected by the low loading obtained for these systems (6-10%) [9]. However, this limited uptake turns out into strong effects on the liposome structure, since progressive formation of multilamellar structures is detected by SAXS experiments, which shows enhanced interactions among bilayers in the presence of porphyrazine molecules [9]. According to the observed SAXS results, in the presence of porphyrazine the fit of the SANS curves could be improved by assuming the presence of at least two populations of different liposomes (Figure 7). In particular, the fit shown in figure 7 was obtained by summing up a 70% fraction of monolamellar vesicles ($R_n$ = 850 Å, $\sigma_R$ = 30 Å, $t_n$ = 37 Å and $\sigma_{tn}$ = 3 Å), and a 30% fraction of bilamellar vesicles with the same structural parameters and interlayer distance 62±2Å. This interlamellar distance was calculated from the corresponding SAXS diagrams [9].

CONCLUSIONS

In this work, cationic, anionic, and zwitterionic liposomes were studied as vectors of a octa(carboranyl)porphyrazine, $H_2$HECASPz. Liposomes were characterized by Light Scattering, Zeta Potential, SAXS, SANS prior and after the insertion of the guest molecules. Two different preparation methods were tested, that is extrusion through polycarbonate membrane of controlled pore sizes (200 nm) and detergent depletion with different helper surfactants. Co-extrusion of lipids and guest molecules resulted in poor loading rates. On the other hand, the detergent depletion method was found to be more suitable for the insertion of $H_2$HECASPz in all the liposomes formulations studied in this work, in spite of the lengthiness and the complications (e.g. need of sonication, use of helper surfactant, higher unpredictability of the liposome final sizes) which are typical of this method. Therefore, we consider that detergent depletion should be preferred with respect to standard preparation procedures in the case of bulky and highly hydrophobic drugs, that are difficult to solubilise in lipid bilayers. This method is also used for the solubilization of large membrane proteins, such as receptors or enzymes, to study their activity inside model membranes [29-30].

The porphyrazine uptake by zwitterionic liposomes was slightly, but appreciably, higher than the uptake by cationic and anionic liposomes. This finding was related to the peculiar interaction previously found to occur between DOPC and DOPE lipids and the tetra-azaporphyrin core [9].

Another interesting point that arose from the results here reported deserves further comments. As mentioned above, the experimental SANS curves were systematically best fitted with slightly lower bilayer thickness values than those obtained in the fittings of SAXS diagrams (Tabs. 1-3). This observed small discrepancy between the SAXS and SANS was in line with the literature on similar systems [21, 23]. It has to be noted, however, that the intensity, in the high-$q$ range was particularly weak ($<10^{-2}$ cm$^{-1}$), because of the low lipid concentration ($10^{-2}$ M) of the liposomes studied in this work. Therefore, no definite answer could be given by SANS fittings at high $q$ in the present cases. We checked with a simple calculation based on eqs. (1-2) and performed according to a homogeneous bilayer model, that a bilayer with hydrated headgroups modelled by three layers of total thickness 40 Å, resulted in a thickness about 35 Å.

To fix ideas, we assumed a much weaker scattering length density in the polar region (of thickness 7.5 Å), due to the hydration, e.g., 6 $10^{10}$ cm$^{-2}$ in the core, 3 $10^{10}$ cm$^{-2}$ in the polar region, and 0 in the solvent. In this case, the exact shape of the high-q oscillations was modified, but as these features were ill-resolved due to the low signal at high-q, the data reported were not sensitive to the shape oscillations. To summarize, the observed discrepancy showed that SANS was less sensitive to the local structure. Moreover, the difference between the thickness values measured from SAXS and SANS was larger in DDLs than in ELs. Since hydration of the lipid heads is known to enhance SAXS/SANS discrepancy in the bilayer thickness values [23], we inferred that a more marked penetration of water molecules in the polar region of the bilayer occurred when liposomes were prepared by detergent depletion. In turn, this enhanced water penetration might be favoured by the pronounced roughness of more polydisperse bilayers.

The overall characterization of loaded liposome vectors, which was done in this work, ensured that after insertion, the liposome structure was not disrupted even if relevant parameters like charge, size and polydispersity were found to change as a function of porphyrazine content and of the employed preparation method.

ACKNOWLEDGMENT. We thank the European Synchrotron Radiation Facility (ESRF, Grenoble) for provision of synchrotron radiation facilities and in particular we would like to thank Dr Emanuela Di Cola for her assistance in using the ID2 beam line. The Laboratoire Léon Brillouin (LLB, Saclay) is also acknowledged for beam time allocation. The Italian Consorzio per le Superfici e per le Grandi Interfasi (CSGI) and Ministero per l'Università e la Ricerca Scientifica are acknowledged for financial support.

FIGURE CAPTIONS

**Figure 1.** Experimental SANS curves of DOTAP/DOPE liposomes prepared by detergent depletion (left) and by extrusion (right). The best fit curves for pure liposomes are also shown.

**Figure 2.** Experimental and fitted SAXS curves of DOTAP/DOPE plain EL and DDL liposomes .

**Figure 3.** Experimental SAXS profiles of plain- and $H_2HECASPz$-loaded cationic DDLs.

**Figure 4.** Experimental SANS curves of the DOPA/DOPE series prepared by detergent depletion and best fit curve of the corresponding pure liposomes.

**Figure 5.** Experimental SAXS curves of the DOPA/DOPE series prepared by extrusion (black points) and by detergent depletion (red points).

**Figure 6.** Experimental SANS profiles of DOPC/DOPE plain and loaded samples prepared by detergent depletion with the detergent sodium cholate (on the left) and $C_{14}TAB$ (on the right). For both cases, the best fit of pure liposome scattering is also showed.

**Figure 7.** Experimental SANS profiles of DOPC/DOPE 100:1 loaded sample prepared by extrusion and the fitting curves of unilamellar and bilamellar liposomes, calculated as described in the text.

**Table 1.** DLS data, ζ and best SAXS and SANS fit parameters for DOTAP/DOPE samples*

|  | Extrusion | | | Detergent depletion with $C_{14}TAB$ | | |
|---|---|---|---|---|---|---|
|  | DOTAP/DOPE | DOTAP/DOPE + $H_2HECASPz$ 200:1 | DOTAP/DOPE + $H_2HECASPz$ 100:1 | DOTAP/DOPE | DOTAP/DOPE + $H_2HECASPz$ 200:1 | DOTAP/DOPE + $H_2HECASPz$ 100:1 |
| Diameter, Å | 1720±50 | 17220±50 | 1800±60 | 1210±30 | 1550±40 | 1550±50 |
| Polyd. Index | 0.02 | 0.05 | 0.07 | 0.30 | 0.40 | 0.53 |
| ζ, mV | +44±3 | +56±3 | +49±2 | +46±5 | +61±5 | +56±5 |
| $\rho_c$, cm$^{-2}$ | 7.80 10$^{10}$ | | | 7.84 10$^{10}$ | 6.23 10$^{10}$ | |
| $\rho_h$, cm$^{-2}$ | 1.31 10$^{11}$ | | | 1.31 10$^{11}$ | 1.41 10$^{11}$ | |
| $t_c$, Å ($\sigma_c$) | 28.0 (1.8) | | | 27.7 (1.5) | 26.4 (0.9) | |
| $t_h$, Å ($\sigma_{t_h}$) | 5.7 | | | 5.0 (0.3) | 7.9 (0.3) | |

| t, Å ($\sigma_t$) | 39.4 (1.8) | | | 37.7 (3.1) | 42.2 (2.3) | |
|---|---|---|---|---|---|---|
| $R_n$, Å ($\sigma_R$) | 850 (10) | 850 (20) | 900 (20) | 600 (50) | 750 (50) | 750 (50) |
| $t_n$, Å ($\sigma_{t_n}$) | 36 (2) | 37 (3) | 37 (3) | 33 (3) | 33 (3) | 34 (3) |

*The liposome diameter in the first row is obtained by DLS measurements. $\rho_c$, $\rho_h$, $t_c$ and $t_h$ scattering length and the thickness of the hydrophobic core and of the polar head, respectively, obtained from the fit of SAXS curves by assuming polydispersity as described by a Gaussian distribution with half-width at half maximum $\sigma_{t_c}$ and $\sigma_{t_h}$. $R_n$ and $t_n$ are the radius and thickness of liposomes, respectively, obtained from the fit of SANS curves by assuming polydispersity as described by a Gaussian distribution with half-width at half maximum $\sigma_R$ and $\sigma_{t_n}$. The n/V used in SANS fitting was calculated the from the $\Sigma$ value obtained from SAXS fitting (4.2 $10^{-4}$ Å$^{-2}$) and from the diameter measured by DLS.

**Table 2.** DLS data, ζ and best SAXS and SANS fit parameters for DOPA/DOPE samples

| | Extrusion | | | Detergent depletion with sodium cholate | | |
|---|---|---|---|---|---|---|
| | DOPA/DOPE | DOPA/DOPE + H$_2$HECASPz 200:1 | DOPA/DOPE + H$_2$HECASPz 100:1 | DOPA/DOPE | DOPA/DOPE + H$_2$HECASPz 200:1 | DOPA/DOPE + H$_2$HECASPz 100:1 |
| Diameter, Å | 1720±50 | 1680±50 | 1650±50 | 1950±60 | 2040±80 | 2180±80 |
| Polyd. Index | 0.11 | 0.09 | 0.04 | 0.16 | 0.08 | 0.10 |
| ζ, mV | -51±5 | -50±5 | -47±8 | -50±4 | -46±3 | -37±4 |
| $\rho_c$, cm$^{-2}$ | 7.85 10$^{10}$ | | | 7.80 10$^{10}$ | 7.5 10$^{10}$ | 7.1 10$^{10}$ |
| $\rho_h$, cm$^{-2}$ | 1.30 10$^{11}$ | | | 1.32 10$^{11}$ | 1.32 10$^{11}$ | 1.37 10$^{11}$ |
| $t_c$, Å ($\sigma_{t_c}$) | 28.5 (1.8) | | | 29.0 (2.2) | 26.0 (2.4) | 25.5 (2.1) |
| $t_h$, Å ($\sigma_{t_h}$) | 6.3 (0.2) | | | 6.4 (0.3) | 7.2 (0.3) | 7.7 (0.3) |
| $t$, Å ($\sigma_t$) | 40.8 (3.1) | | | 40.8 (4) | 40.4 (4.4) | 40.4 (4.1) |
| $R_n$, Å ($\sigma_R$) | 850 (10) | 850 (10) | 850 (10) | 950 (10) | 1000 (10) | 1000 (20) |
| $t_n$ Å ($\sigma_{t_n}$) | 34 (2) | 34 (3) | 34 (3) | 33 (2) | 33 (3) | 34 (3) |

\* See Table 1 for the meaning of the symbols. $\Sigma = 4.6 \cdot 10^{-4}$ Å$^{-2}$, obtained from SAXS fitting

**Table 3.** DLS data and best SANS fit parameters for DOPC/DOPE liposomes prepared by detergent depletion.

| | Detergent depletion with sodium cholate | | | Detergent depletion with C$_{14}$TAB | | |
|---|---|---|---|---|---|---|
| | DOPC/DOPE | DOPC/DOPE + H$_2$HECASPz 200:1 | DOPC/DOPE + H$_2$HECASPz 100:1 | DOPC/DOPE | DOPC/DOPE + H$_2$HECASPz 200:1 | DOPC/DOPE + H$_2$HECASPz 100:1 |
| Diameter, Å | 1570±50 | 1730±60 | 1810±60 | 1090±30 | 1390±40 | 1570±50 |
| Polyd. Index | 0.09 | 0.19 | 0.14 | 0.39 | 0.38 | 0.48 |
| $R_n$, Å | 600 (20) | 850 (30) | 850 (30) | 550 (80) | 700 (80) | 800 (30) |

| ($\sigma_R$) | | | | | | |
|---|---|---|---|---|---|---|
| $t_n$, Å ($\sigma_{t_n}$) | 34 (3) | 34 (3) | 35 (3) | 34 (3) | 35 (3) | 35 (4) |

\* See Table 1 for the meaning of the symbols. $\Sigma = 4.9 \; 10^{-4}$ Å$^{-2}$, obtained from SAXS fitting

REFERENCES

(1) Heurtault, B.; Saulnier, P.; Pech, B.; Proust, J.-E.; Benoit, J.-P. *Biomaterials* **2003**, *24*, 4283-4300.

(2) Charrois, G. J. R.; Allen, T. M. *Biochim. Biophys. Acta* **2004**, *1663*, 167-177.

(3) Gregoriadis, G. *FEBS Letters* **1973**, *36*, 292-296.

(4) Davidson, R. N.; Croft, S. L.; Scott, A.; Maini, M.; Moody, A. H.; Bryceson, A. D. *Lancet* **1991**, *337*, 1061-1062.

(5) Guaglianone, P.; Chan, K.; Delaflor-Weiss, E.; Hanisch, R.; Jeffers, D.; Sharma, D.; Muggia, F. *Invest. New Drugs* **1994**, *12*, 103-112.

(6) Torchilin, V. P. *Nature Rev.* **2005**, *4*, 145-160.

(7) Safinya, C. R. *Curr. Opin. Structural Biol.* **2001**, *11*, 440-448.

(8) Meidan, V. M.; Glezer, J.; Amariglio, N.; Choen, J. S.; Barenholz, Y. *Biochim. Biophys. Acta* **2001**, *1568*, 177-182.

(9) Ristori, S.; Salvati, A.; Martini, G.; Spalla, O.; Pietrangeli, D.; Rosa, A.; Ricciardi, G. *J. Am.Chem. Soc.* **2007**, *129*, 2728-2729.

(10) Monkkonen, J.; Urtti, A. *Adv. Drug Delivery Rev.* **1998**, *34*, 37-49.

(11) Oberle, V.; Bakowsky, U.; Zuhorn, I. S.; Hoekstra, D. *Biophys. J.* **2000**, *79*, 1447-1454.

(12) Hirsch-Lerner, D.; Zhang, M.; Eliyahu, H.; Ferrari, M. E.; Wheeler, C. J.; Barenholz, Y. *Biochim. Biophys. Acta* **2005**, *1714*, 71-84.

(13) Zuidam, N. J.; Barenholz, Y. *Biochim Biophys Acta* **1997**, *1329*, 211-222.

(14) New, R. R. C., Liposomes: a practical approach, IRL Press, New York, 1990.

(15) Brunner, J.; Skrabal, P.; Hausser, H. *Biochim. Biophys. Acta* **1976**, *455*, 322-331.

(16) Ciani, L.; Ristori, S.; Salvati, A.; Calamai, L.; Martini, G. *Biochim. Biophys. Acta* **2004**, *1664*, 70-79.

(17) Koppel, D. E. *J. Chem. Phys.* **1972**, *57*, 4814-4820.

(18) Provencher, S. W. *Comp. Phys. Comm.* **1982**, *27*, 213-227.

(19) Pedersen, J. S.; Posselt, D.; Mortensen, K. *J. Appl. Cryst.* **1990**, *23*, 321-333.

(20) Lairez, D. *J Phys IV* **1999**, *9*, 67-81.

(21) Kiselev, M. A.; Zemlyanaya, E. V.; Aswal, V. K.; Neubert, R. H. H. *Eur. Biophys. J.* **2006**, *35*, 477-493.

(22) Wiener, M. C.; White, S. H. *Biophys. J.* **1991**, *59*, 162-173.

(23) Armen, R. S.; Uitto, O. D.; Feller, S. E. *Biophys. J.* **1998**, *75*, 734-744.

(24) Bosecke, P.; Diat, O. *J. Appl. Cryst.* **1997**, *30*, 867-871.

(25) Lindner, P.; Zemb, T., Neutron, X-rays and Light. Scattering Methods Applied to Soft Condensed Matter, North Holland Press, Amsterdam, 2002.

(26) Cantù, L.; Corti, M.; Del Favero, E.; Dubois, M.; Zemb, T. N. *J. Phys. Chem. B* **1998**, *102*, 5737-5743.

(27) Ristori, S.; Oberdisse, J.; Grillo, I.; Donati, A.; Spalla, O. *Biophys. J.* **2005**, *88*, 535-547.

(28) Allen, T. M.; Romans, A. Y.; Kercret, H.; Segrest, J. P. *Biochim. Biophys. Acta* **1980**, *601*, 328-342.

(29) Giustini, M.; Castelli, F.; Husu, I.; Giomini, M.; Mallardi, A.; Palazzo, G. *J. Phys. Chem.* **2005**, *109*, 21187-21196.

(30) Palazzo, G.; Mallardi, A.; Giustini, M.; Berti, D.; Venturoli, G. *Biophys. J.* **2000**, *79*, 1171–1179.

Figure 1

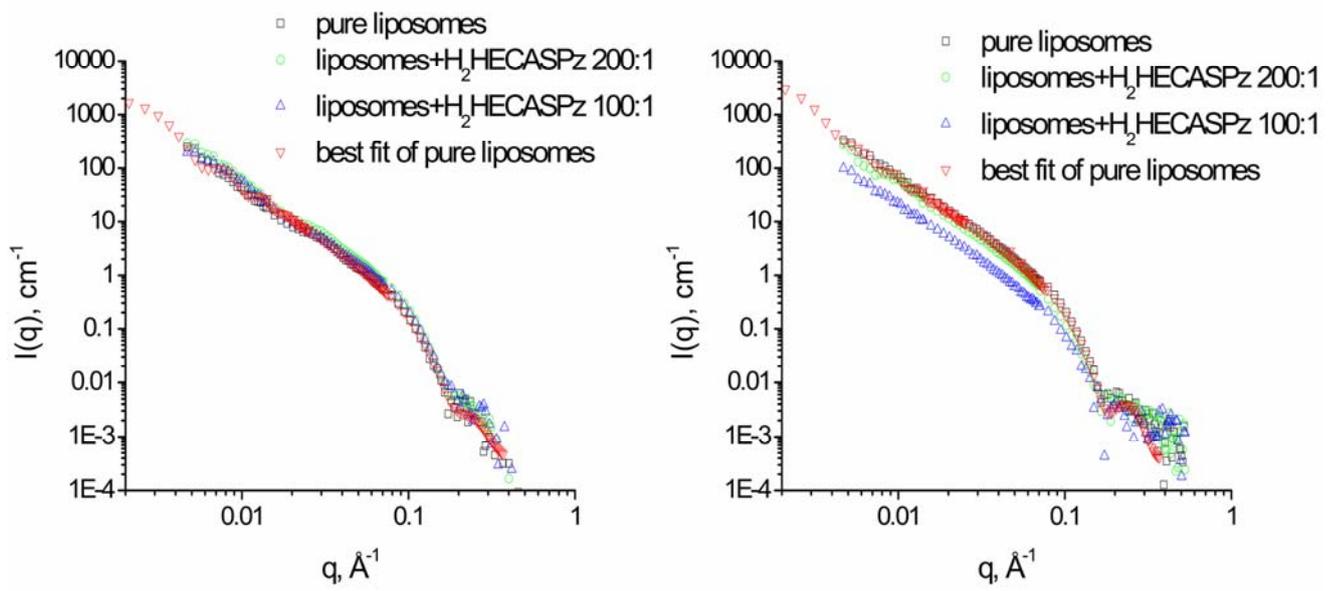

Figure 2

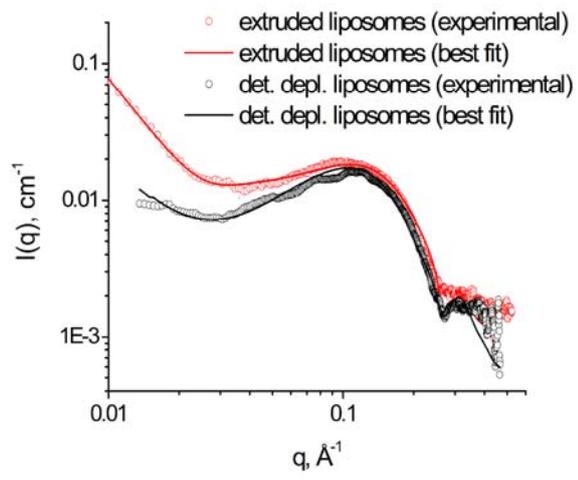

Figure 3

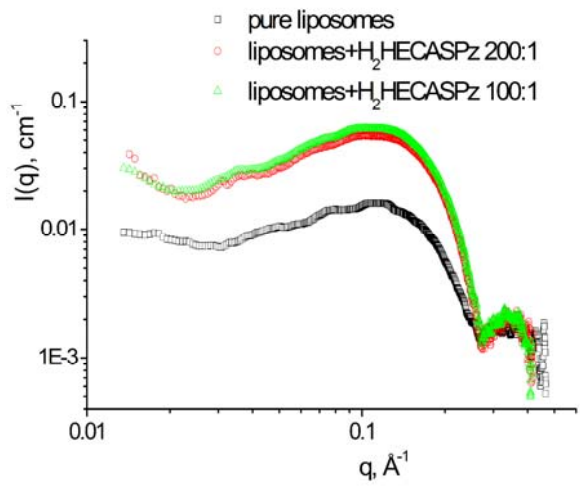

Figure 4

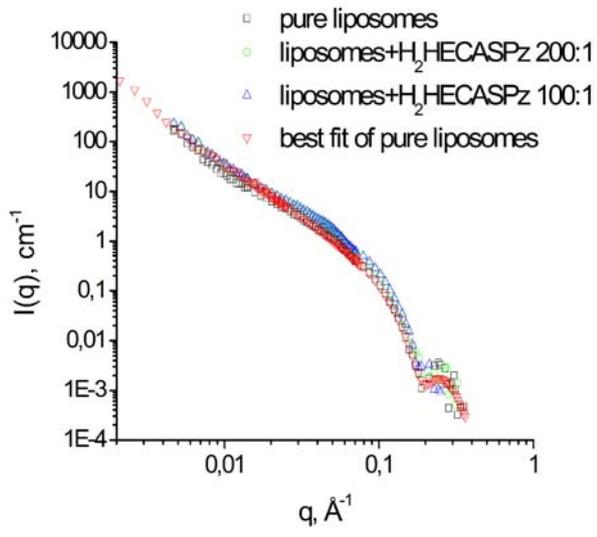

Figure 5

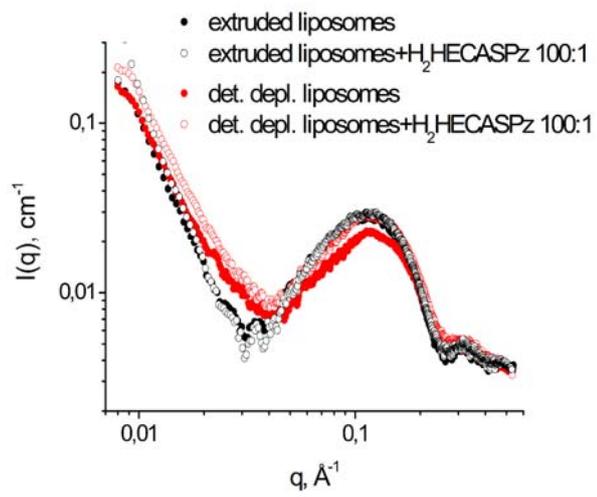

Figure 6

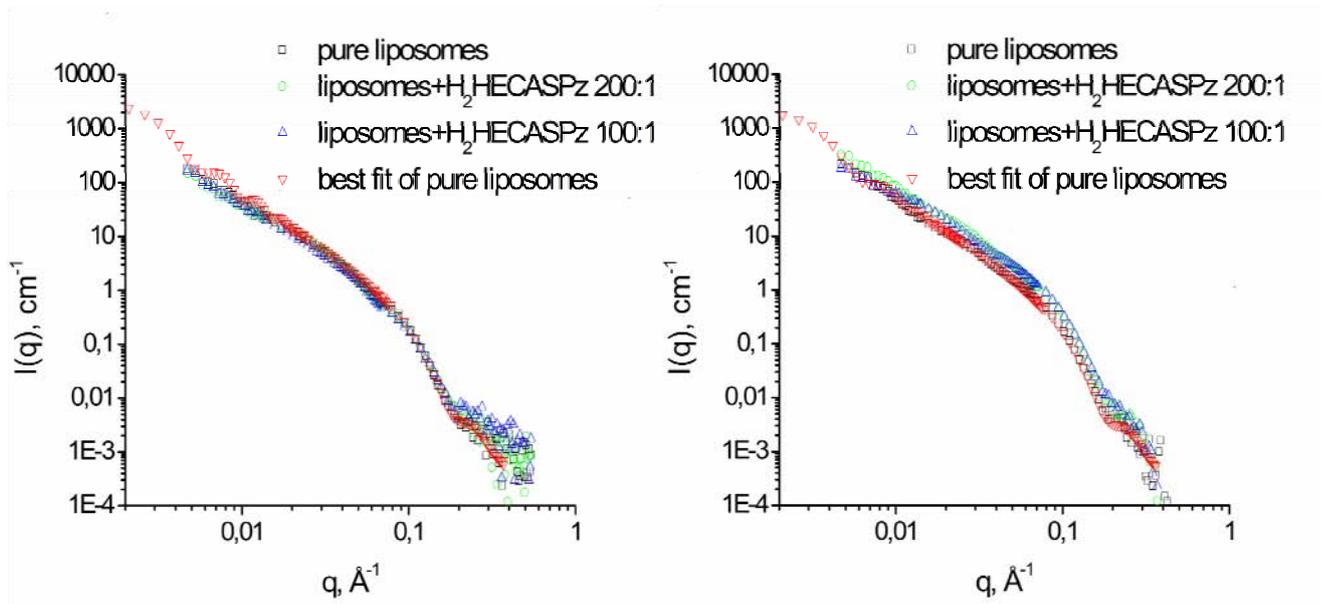

Figure 7

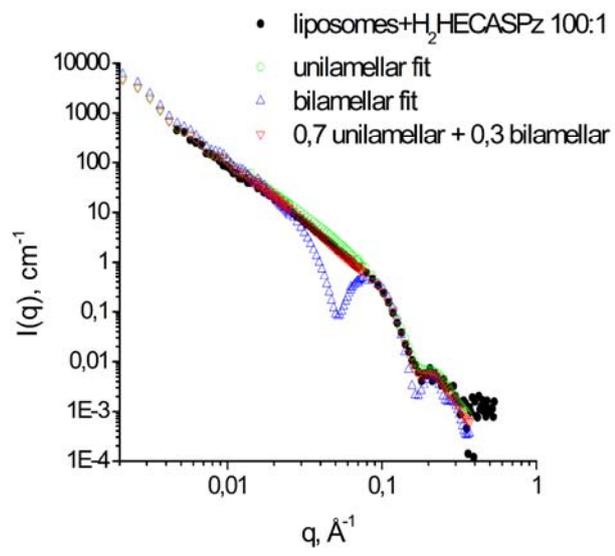